\documentclass[11pt,final]{article}
\usepackage{amsmath, amssymb, amsthm}
\usepackage{subfigure}
\usepackage{graphicx}

\title{Small-$Q^2$ extension of DGLAP-constrained Regge residues}
\author{G. Soyez\footnote{e-mail: g.soyez@ulg.ac.be}}

\newcommand{\xgrv}{x_{\text{Regge}}}
\newcommand{\qmin}{Q^2_0}

\begin{document}

\maketitle

\begin{abstract}
In a previous paper, we have shown that it was possible to use the DGLAP evolution equation to constrain the high-$Q^2$ ($Q^2 \ge 10$ GeV$^2$) behaviour of the residues of a high-energy Regge model, and we applied the developed method to the triple-pole pomeron model. We show here that one can obtain a description of the low-$Q^2$ $\gamma^{(*)} p$ data matching the high-$Q^2$ results at $Q^2 = 10$ GeV$^2$.
\end{abstract}

We know that one can use Regge theory \cite{books} to describe high-energy hadronic interactions. Particularly, using a triple-pole pomeron model \cite{Cudell:2001ii,Desgrolard:2001bu,Cudell:2002xe}, one can reproduce the hadronic total cross-sections, the $\gamma p$ and $\gamma\gamma$ cross-sections, and also the proton and photon structure functions $F_2^p$ and $F_2^\gamma$. In the latter case, one must point out that Regge theory is applied at all values of $Q^2$. 

On the other hand, it is well known that the high-$Q^2$ behaviour of the proton structure function can be reproduced using the DGLAP evolution equation \cite{DGLAP}. Therefore, we would like to find a model compatible both with Regge theory and with DGLAP evolution at high $Q^2$. We have shown \cite{Soyez:2003sr} that it is possible to extract the behaviour of the triple-pole pomeron residues at high $Q^2$ from DGLAP evolution. In such an analysis, we need information not only on $F_2$ but also on parton distributions. One easily shows that the minimal number of quark distributions needed to reproduce $F_2^p$ is 2: one flavour-non-singlet distribution
\[
T(x,Q^2) = x\left[(u^++c^++t^+)-(d^++s^++b^+)\right],
\]
with $q^+ = q+\bar{q}$, evolving alone with $xP_{qq}$ as splitting function, and one flavour-singlet distribution
\[
\Sigma(x,Q^2) = x\left[(u^++c^++t^+)+(d^++s^++b^+)\right],
\]
coupled with the gluon distribution $xg(x,Q^2)$ and evolving with the full splitting matrix. Before going into the main subject of this paper, we shall summarise the techniques developed in this previour paper \cite{Soyez:2003sr} and show how we can extend the results down to $Q^2=0$.

First of all, given that $F_2$ can be parametrised at small $x$ by a $\log^2(1/x)$ term, we have parametrised the quark content of the proton in the most natural way {\em i.e.} using a triple-pole pomeron term and an $f/a_2$ reggeon terms. After a few manipulations, we end up with the following functions
\begin{eqnarray}\label{eq:initparam}
T(x,Q_0^2) & = & d_T^* x^\eta (1-x)^{b_2},\nonumber\\
\Sigma(x,Q_0^2) & = & a_\Sigma \log^2(1/x) + b_\Sigma \log(1/x) + c_\Sigma^* (1-x)^{b_1}\nonumber\\
                & + & d_\Sigma x^\eta(1-x)^{b_2}\nonumber\\
xg(x,Q_0^2) & = & a_G \log^2(1/x) + b_G \log(1/x) + c_G^* (1-x)^{b_1}.
\end{eqnarray}
Since Regge theory does not extend up to $x=1$, we used the GRV parametrisation for $x\ge \xgrv = 0.15$ and imposed that our distributions match GRV's at $x=\xgrv$. This requirement constrains the parameters marked with a superscript $^*$ in eq. \eqref{eq:initparam}. Thus, the 7 parameters $a_\Sigma$, $b_\Sigma$, $d_\Sigma$, $a_G$, $b_G$, $b_1$ and $b_2$ need to be extracted from DGLAP evolution.

Since DGLAP evolution generates an essential singularity in the complex-$j$ plane at $j=1$, the only place where we can use the Regge model is in the initial distributions at $Q^2 = Q_0^2$. In such a case, we shall not worry about the presence of an essential singularity for $Q^2\neq Q_0^2$ and consider the result of DGLAP evolution as a numerical approximation to a triple-pole pomeron. One can therefore extract the residues of the Regge model at high $Q^2$ using the following method:
\begin{enumerate}
\item\label{s21} choose an initial scale $Q_0^2$,
\item\label{s22} choose a value for the parameters in the initial distribution,
\item\label{s23} compute the parton distributions for $Q_0^2 \le Q^2  \le Q_{\text{max}}^2$ using forward DGLAP evolution and for $Q_{\text{min}}^2 \le Q^2  \le Q_0^2$ using backward DGLAP evolution,
\item\label{s24} repeat \ref{s22} and \ref{s23} until the value of the parameters reproducing the $F_2$ data for $Q^2>Q_{\text{min}}^2$ and $x\le \xgrv$ is found.
\item This gives the residues at the scale $Q_0^2$ and steps \ref{s21} to \ref{s24} are repeated in order to obtain the residues at all $Q^2$ values.
\end{enumerate}

We have applied this method to the parametrisation \eqref{eq:initparam} within the domain
\begin{equation}\label{eq:domain}
\begin{cases}
10 \le Q^2 \le 1000\:\text{GeV}^2, \\
\cos(\theta_t) = \frac{\sqrt{Q^2}}{2xm_p} \ge \frac{49\:\text{GeV}^2}{2m_p^2},
\end{cases}
\end{equation}
ensuring that both Regge theory and DGLAP evolution can be applied, and required\footnote{This limit is only effective at large $Q^2$.} $x<0.15$. Using the residues of the triple-pole pomeron obtained in this way, we have a description of $F_2^p$ for $Q^2\ge 10$ GeV$^2$ with a $\chi^2/nop$ of 1.02 for 560 experimental points. 

Since the method explained here gives us the Regge residues at large scales, one may ask if it is possible to extend the results down to $Q^2=0$. The main problem here is that, instead of using $x$ and $Q^2$, we must use $\nu$ and $Q^2$ if we want to obtain a relevant expression for the total cross section. Of course, we shall only extend the $F_2^p$ predictions instead of the parton distributions $T$ and $\Sigma$.

As a starting point, we shall not consider the powers of $(1-x)$ since, at low $Q^2$, there are no point inside the Regge domain beyond $x=0.003$, which means that it is just a correction of a few percents. At low $Q^2$, we require that $F_2$ has the same form as used in \cite{Cudell:2001ii}
\begin{equation}
F_2(\nu, Q^2) = \frac{Q^2}{4\pi^2\alpha_e} \left\{A(Q^2)\left[\log(2\nu)-B(Q^2)\right]^2+C(Q^2)+D(Q^2)(2\nu)^{-\eta}\right\}.
\end{equation}
The total $\gamma p$ cross-section is then
\begin{equation}
\sigma_{\gamma p} = A(0)\left[\log(s)-B(0)\right]^2+C(0)+D(0)s^{-\eta}.
\end{equation}
At $Q^2=Q_0^2$, the form factors $A$, $B$, $C$ and $D$ are related to the parametrisation \eqref{eq:initparam} by the relations
\begin{eqnarray}\label{eq:link}
A(\qmin) & = & \frac{4\pi^2\alpha_e}{\qmin}a_0,\nonumber \\
B(\qmin) & = & \log(\qmin)-\frac{b_0}{2a_0},\nonumber\\[-3mm]
&&\\[-3mm]
C(\qmin) & = & \frac{4\pi^2\alpha_e}{\qmin}\left(c_0-\frac{b_0^2}{4a_0}\right),\nonumber\\
D(\qmin) & = & \frac{4\pi^2\alpha_e}{\qmin}d_0 (\qmin)^\eta.\nonumber
\end{eqnarray}
where the subscript $_0$ to refer to the form factors obtained at $Q^2=Q_0^2$ from DGLAP evolution.

At small $Q^2$, the unknown functions $A$, $B$, $C$ and $D(Q^2)$ are parametrised in the same way as in \cite{Cudell:2001ii}
\begin{eqnarray}\label{eq:param}
A(Q^2) & = & A_a\left(\frac{Q_a^2}{Q^2+Q_a^2}\right)^{\varepsilon_a},\nonumber\\
B(Q^2) & = & A_b\left(\frac{Q^2}{Q^2+Q_b^2}\right)^{\varepsilon_b}+A_b',\nonumber\\[-3mm]
&&\\[-3mm]
C(Q^2) & = & A_c\left(\frac{Q_c^2}{Q^2+Q_c^2}\right)^{\varepsilon_c},\nonumber\\
D(Q^2) & = & A_d\left(\frac{Q_d^2}{Q^2+Q_d^2}\right)^{\varepsilon_d}.\nonumber
\end{eqnarray}

If we use the relations \eqref{eq:link} to fix the parameters $A_a$, $A_b'$, $A_c$ and $A_d$ in \eqref{eq:param}, we find the final form of the small-$Q^2$ form factors:
\begin{eqnarray}
  A(Q^2) & = & \frac{4\pi^2\alpha_e}{\qmin} a_0 \left(\frac{\qmin+Q^2}{Q_a^2+Q^2}\right)^{\varepsilon_a},\nonumber\\
  B(Q^2) & = & \log(\qmin) -\frac{b_0}{2a_0}+A_b\left\lbrack\left(\frac{Q^2}{Q_b^2+Q^2}\right)^{\varepsilon_b}-\left(\frac{\qmin}{Q_b^2+\qmin}\right)^{\varepsilon_b}\right\rbrack,\nonumber\\[-3mm]
&&\\[-3mm]
  C(Q^2) & = & \frac{4\pi^2\alpha_e}{\qmin} \left(c_0-\frac{b_0^2}{4a_0}\right) \left(\frac{\qmin+Q^2}{Q_c^2+Q^2}\right)^{\varepsilon_c},\nonumber\\
  D(Q^2) & = & \frac{4\pi^2\alpha_e}{\qmin} d_0 (\qmin)^\eta \left(\frac{\qmin+Q^2}{Q_d^2+Q^2}\right)^{\varepsilon_d}.\nonumber
\end{eqnarray}

\begin{table}
\begin{center}
\begin{tabular}{|c||c|c|}
\hline
   Parameter    &   value  & error   \\
\hline
\hline
$A_b$           & 69.151    & 0.055    \\
\hline
$Q^2_a$         & 25.099    & 0.088    \\
\hline
$Q^2_b$         & 4.943     & 0.086   \\
\hline
$Q^2_c$         &  0.002468 & 0.000042 \\
\hline
$Q^2_d$         &  0.01292  & 0.00074    \\
\hline
$\varepsilon_a$ &  1.5745   & 0.0046   \\
\hline
$\varepsilon_b$ &  0.08370  & 0.00052   \\
\hline
$\varepsilon_c$ &  0.92266  & 0.00019  \\
\hline
$\varepsilon_d$ &  0.3336   & 0.0029  \\
\hline
\end{tabular}
\end{center}
\caption{Values of the parameters for the low-$Q^2$ fit ($0\le Q^2\le \qmin$).}\label{tab:toq0}
\end{table}

If we now want to reinsert the large-$x$ corrections, we need to multiply $c$ and $d$ by some power of $(1-x)$. This gives
\begin{eqnarray*}
\frac{4\pi^2\alpha_e}{Q^2} F_2(x, Q^2) 
 & = & A(Q^2)\log(1/x)\left\{\log(1/x)+2\left[\log(Q^2)-B(Q^2)\right]\right\}\\
 & + & \left\{A(Q^2)\left[\log(Q^2)-B(Q^2)\right]^2+C(Q^2) \right\}(1-x)^{b_1}\\
 & + & D(Q^2) \left(\frac{Q^2}{x}\right)^{-\eta}(1-x)^{b_2}.
\end{eqnarray*}

These large-$x$ corrections do not modify the expression of the total cross section since, when $Q^2\to 0$
\[
1-x = 1-\frac{2\nu}{Q^2} \to 1.
\]
Moreover, since the large-$x$ corrections are only a few percents effects, we shall keep the exponents $b_1$ and $b_2$ constant and equal to their value at $Q^2=Q_0^2$.

\begin{figure}[!ht]
\begin{center}
\subfigure{\includegraphics[scale=0.8]{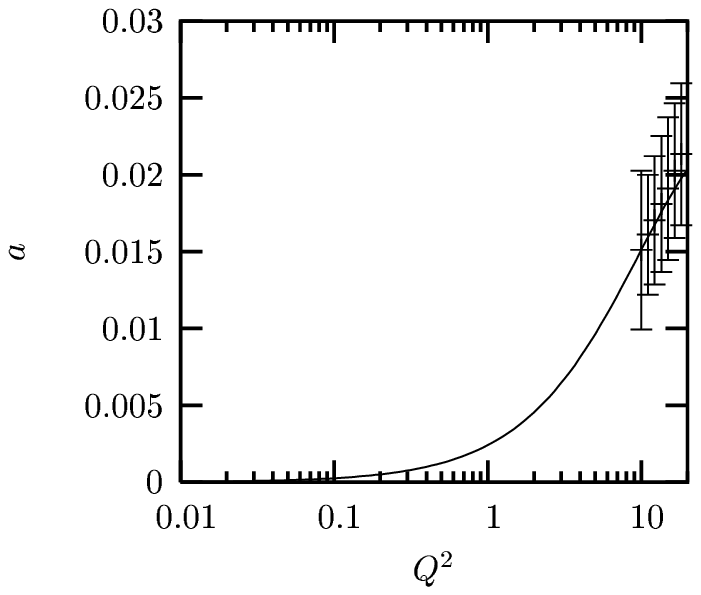}}
\subfigure{\includegraphics[scale=0.8]{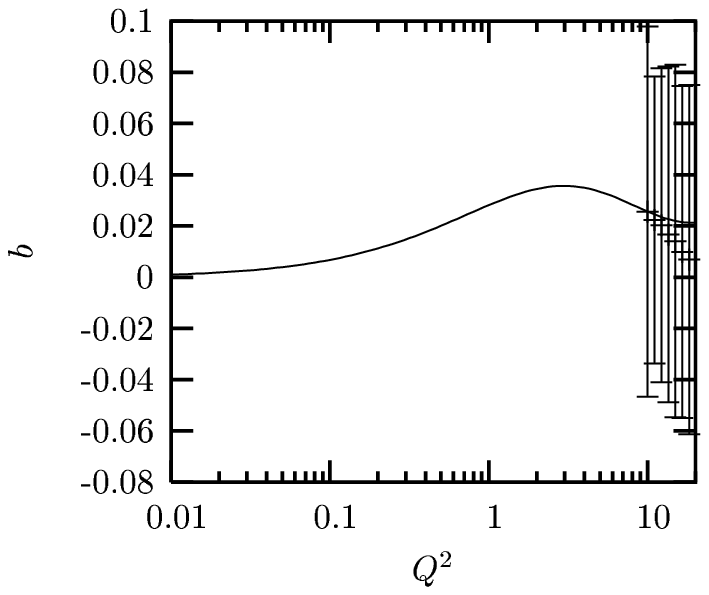}}
\subfigure{\includegraphics[scale=0.8]{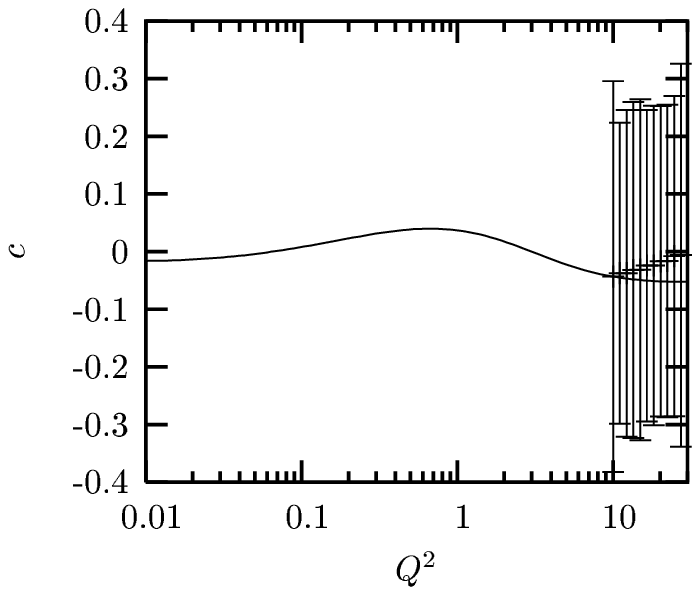}}
\subfigure{\includegraphics[scale=0.8]{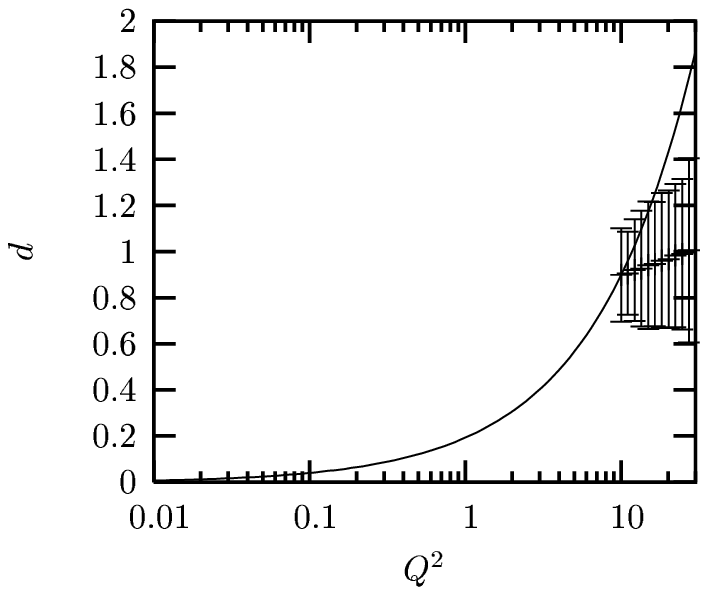}}
\end{center}
\caption{Regge theory predictions for the form factors at small values of $Q^2$. The lines show the analytical curve for $0\le Q^2\le 10$ GeV$^2$ and the points are the results obtained in \cite{Soyez:2003sr} from DGLAP evolution.}\label{fig:toq0-form}
\end{figure}

\begin{table}
\begin{center}
\begin{tabular}{|l||c|c|c|}
\hline
Experiment & $n$ & $\chi^2$ & $\chi^2/n$ \\
\hline
\hline
  E665     &  69 &   59.811 & 0.867 \\
\hline
  H1       &  99 &  104.924 & 1.060 \\
\hline
  NMC      &  37 &   28.392 & 0.767 \\
\hline
  ZEUS     & 216 &  201.790 & 0.934 \\
\hline
\hline
  $F_2^p$  & 421 &  394.916 & 0.938 \\
\hline
$\sigma_
{\gamma p}$&  30 &   17.171 & 0.572 \\
\hline
\hline
Total      & 451 &  412.086 & 0.914 \\
\hline
\end{tabular}
\end{center}
\caption{$\chi^2$ resulting from the small-$Q^2$ Regge fit. The results are given for all $F_2^p$ experiments and for the total cross-section.}\label{tab:toq0-chi2}
\end{table}

\begin{figure}[ht]
\begin{center}
\includegraphics{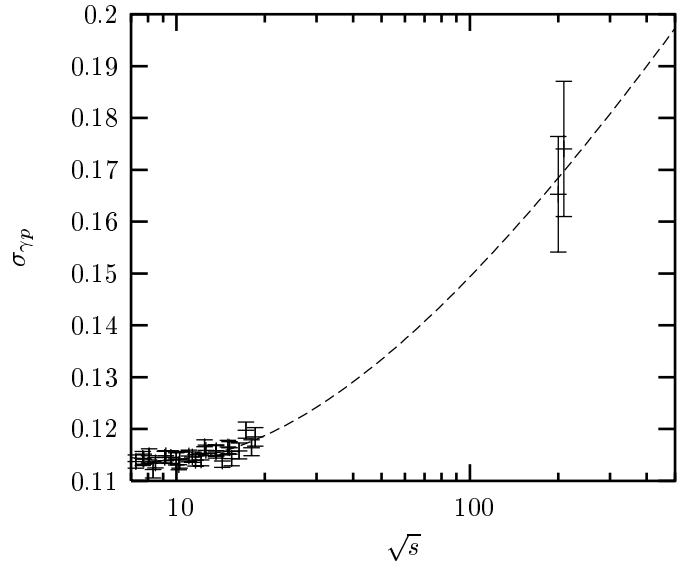}
\end{center}
\caption{Fit for the total $\gamma p$ cross-section.}\label{fig:toq0-sig}
\end{figure}

\begin{figure}[ht]
\hspace{-2.0cm}\includegraphics{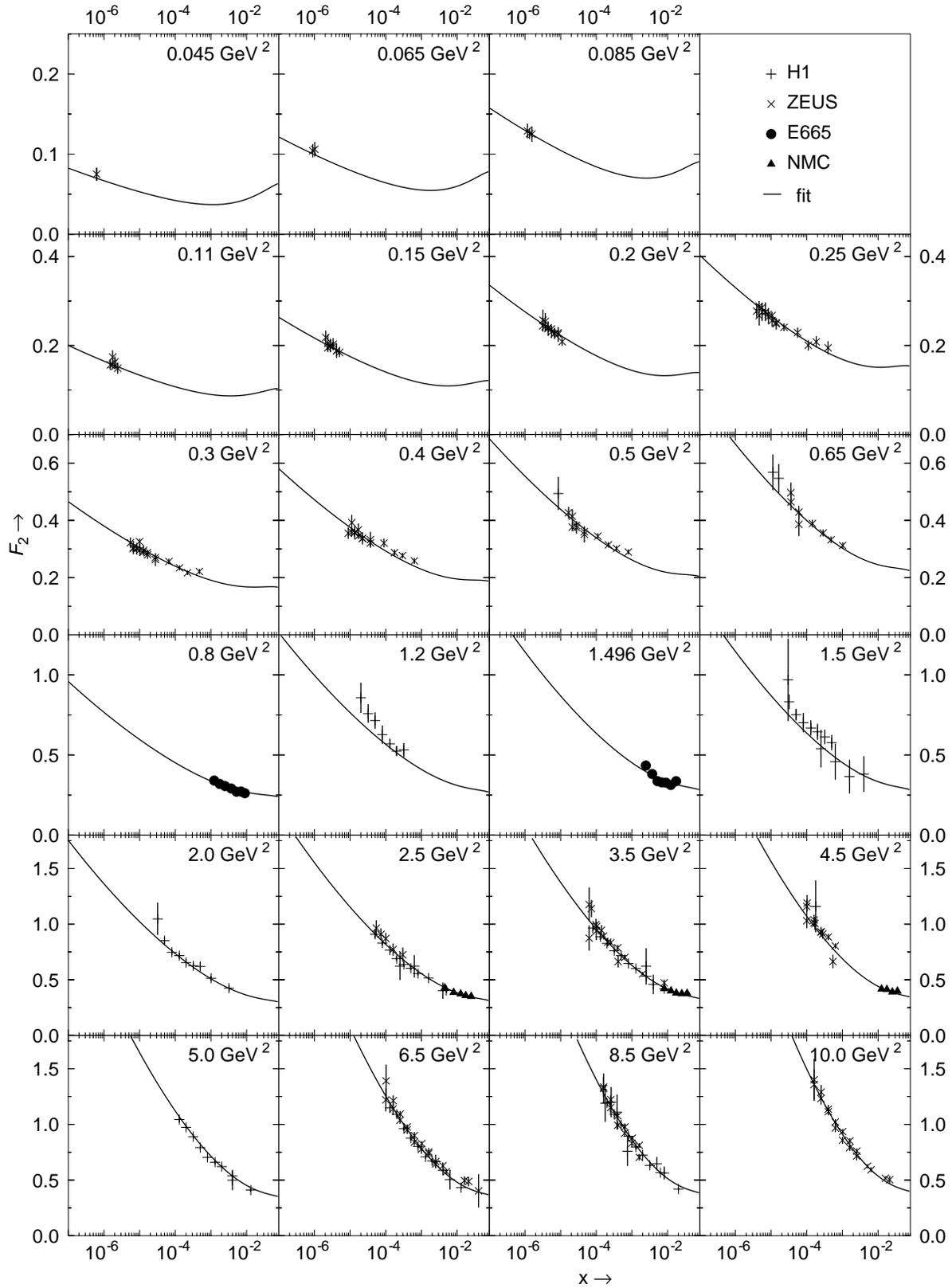}
\caption{Fit for the $F_2^p$ at low $Q^2$. Only the most populated $Q^2$ bins are shown.}\label{fig:toq0-f2}
\end{figure}

Now, we may adjust the parameters in the form factors by fitting $F_2^p$ in the Regge domain 
\begin{equation}\label{eq:domain2}
\begin{cases}
\nu \ge 49\:\text{GeV}^2,\\
\cos(\theta_t) = \frac{\sqrt{Q^2}}{2xm_p} \ge \frac{49\:\text{GeV}^2}{2m_p^2},\\
Q^2 \le 10\:\text{GeV}^2,
\end{cases}
\end{equation}
together with the total cross-section for $\sqrt{s}\ge 7$ GeV. The resulting parameters are presented in Table \ref{tab:toq0} and the form factor are plotted in Figure \ref{fig:toq0-form}. As we can see from Table \ref{tab:toq0-chi2} and from Figures \ref{fig:toq0-sig} and \ref{fig:toq0-f2}, this gives a very good extension in the soft region (see Table \ref{tab:toq0-chi2}).

To conclude, we have seen that, using a triple-pole-pomeron model, one can obtain a description of the $\gamma^{(*)}p$ interactions at all values of $Q^2$ compatible with the DGLAP equation at large $Q^2$. It should be interesting, in the future, to test this method with other Regge models and to see if the results are compatible with the $t$-channel unitarity relations obtained in \cite{Cudell:2002ej} and if they can give useful information on how to link perturbative and non-perturbative QCD.

\begin{center}{\bf Acknowledgments}\end{center}
I would like to thanks J.-R. Cudell for useful discussions. This work is supported by the National Fund for Scientific Research (FNRS), Belgium.

\end{document}